# Ultra Violet Imaging Telescope (UVIT) on ASTROSAT

Amit Kumar[1], S. K. Ghosh[2], J. Hutchings[3], P.U.Kamath[1], S.Kathiravan[1], P.K.Mahesh[1], J. Murthy[1], Nagbhushana.S[1], A.K. Pati[1], M. N.Rao[1], N. K. Rao[1], S.Sriram[1], and S.N.Tandon[1,4]

1. Indian Institute of Astrophysics, Bangalore
2. National Centre for Radio Astrophysics, TIFR, Pune
3. NRC Herzberg Institute of Astrophysics, Canada
4. Inter University Centre for Astronomy and Astrophysics, Pune

## ABSTRACT

Ultra Violet Imaging Telescope on ASTROSAT Satellite mission is a suite of Far Ultra Violet (FUV; 130 – 180 nm), Near Ultra Violet (NUV; 200 – 300 nm) and Visible band (VIS; 320–550nm) imagers. ASTROSAT is the first multi wavelength mission of INDIA. UVIT will image the selected regions of the sky simultaneously in three channels & observe young stars, galaxies, bright UV Sources. FOV in each of the 3 channels is ~ 28 arc-minute. Targeted angular resolution in the resulting UV images is better than 1.8 arc-second (better than 2.0 arc-second for the visible channel). Two identical co-aligned telescopes (T1, T2) of Ritchey-Chretien configuration (Primary mirror of ~375 mm diameter) collect celestial radiation and feed to the detector system via a selectable filter on a filter wheel mechanism; gratings are available in filter wheels of FUV and NUV channels for slit-less low resolution spectroscopy. The detector system for each of the 3 channels is generically identical. One of the telescopes images in the FUV channel, while the other images in NUV and VIS channels. Images from VIS channel are also used for measuring drift for reconstruction of images on ground through shift and add algorithm, and to reconstruct absolute aspect of the images. Adequate baffling has been provided for reducing scattered background from the Sun, earth albedo and other bright objects. One time open-able mechanical cover on each telescope also works as a Sun-shield after deployment. We are presenting here the overall (mechanical, optical and electrical) design of the payload.

## 1. INTRODUCTION

ASTROSAT mission of Indian Space Research Organisation (ISRO) is planned for multi-wavelength astronomical observations covering the range from the visible band to hard X-rays, and is expected to be launched by ISRO launch vehicle PSLV in the year 2013. There are four independent instruments for observing in the X-ray bands and a Ultra Violet Imaging Telescope (UVIT) for observing in visible and ultraviolet bands. UVIT is primarily an imaging instrument which makes images with FWHM < 1.8" (for comparision images from GALEX[1] have FWHM of 4 to 5") over a field of ~ 28' simultaneously in three channels, namely VIS (320-550 nm), NUV (200-300 nm), and FUV (130-180 nm). (For an early concept of UVIT please refer to reference 2)

After a brief presentation of the scientific objectives, we present details of the design of the payload and its implementation, as well as the expected performance as inferred from the tests on ground.

## 2. SCIENTIFIC GOALS AND OBJECTIVES

One of the key objectives of UVIT is to observe time variations, in the visible and ultraviolet bands, of X-ray sources simultaneously with the X-ray telescopes on ASTROSAT. In addition a large variety of other sources would be observed. Some of the examples are: globular-clusters, planetary nebulae, high resolution studies of galaxy morphology in the UV, star-formation in interacting galaxies, star-formation history of the universe, and deep UV survey of selected parts of the sky (more details, see reference 3).

*amits@iiap.res.in; phone 91 80 22541253; fax 91 80 25534043; iiap.res.in

# 3. INSTRUMENTATION

## 3.1 General

The payload (UVIT) is configured as a twin telescope: one of these makes images in FUV and the other makes images in NUV and VIS; the radiation is divided between NUV and VIS channels by a dichroic beam splitter. Each of the two telescopes is a RC configuration, with an aperture of ~375mm and a focal length of ~ 4750 mm, see Figure 1. The images from VIS channel are also used to find aspect of UVIT about once per second. For selection of a band within each of the three channels a set of filters is mounted on a wheel; this wheel also carries a blind to block radiation. See section 3.2 for details of the optical layout. The wheels for NUV and FUV channels also carry gratings to provide low resolution (~ 100) slit-less spectroscopy. Photon counting imaging detectors are used in all the three channels to get a resolution of ~ 1.8" FWHM. The detectors can also be used with a low gain (called integration mode), but in this case individual photons are not detected and the spatial resolution is ~ 3". As the satellite is not stabilized to better than 10", it is also required that short exposures are taken and are integrated through a shift and add algorithm on ground: the shift is found by comparing successive images from VIS channel taken every second or so. The success of this algorithm depends on the absence of any jitter > 0.3" rms in attitude of the satellite (either due to some internal motions of any P/L etc. or otherwise), and a drift free relative aspect of the three channels over periods of ~ 1000 s: a duration which is large enough to collects enough photons from sources in the UV images.

Those parts of the telescopes which define locations of the optical elements are mostly made of Invar36, and the other parts are made of Aluminium alloy. The two telescopes are mounted on a cone-like structure made of Titanium, which is attached to central cylinder of the spacecraft. Active thermal control is used to keep temperature of the optical tube within ±3°C. In order to carry out 'de-contamination' of the important optical components, an independent set of dedicated heaters have been provided which can be turned ON if required for a desired duration.

A cylindrical baffle extends over each of the telescopes for attenuating the radiation from off-axis sources. With these baffles light reaching the detector from sources at $45^0$ from the axis is attenuated by a factor $10^9$; with such attenuation light reaching the detector from full Moon at $45^0$ from the axis is less than the average sky background. In addition to these baffles, the doors act as sun shades as long as Sun is at > $45^0$ from the axis. In order to avoid contamination of the optics due to ultraviolet assisted reactions, bright earth-limb is kept away from the axis by >$12^0$ and the sun is kept behind the sun-shield at all times even if UVIT is not observing. The geo-coronal lines are very strong in day time, and a significant amount of solar radiation could be scattered by the other instruments into the baffles. Therefore, the nominal observation period is restricted to the night time (In special cases, observations in day time could be considered).

As ultraviolet optics is very sensitive to molecular contaminations, an elaborate plan has been implemented to avoid any such contamination during the assembly and testing phases.

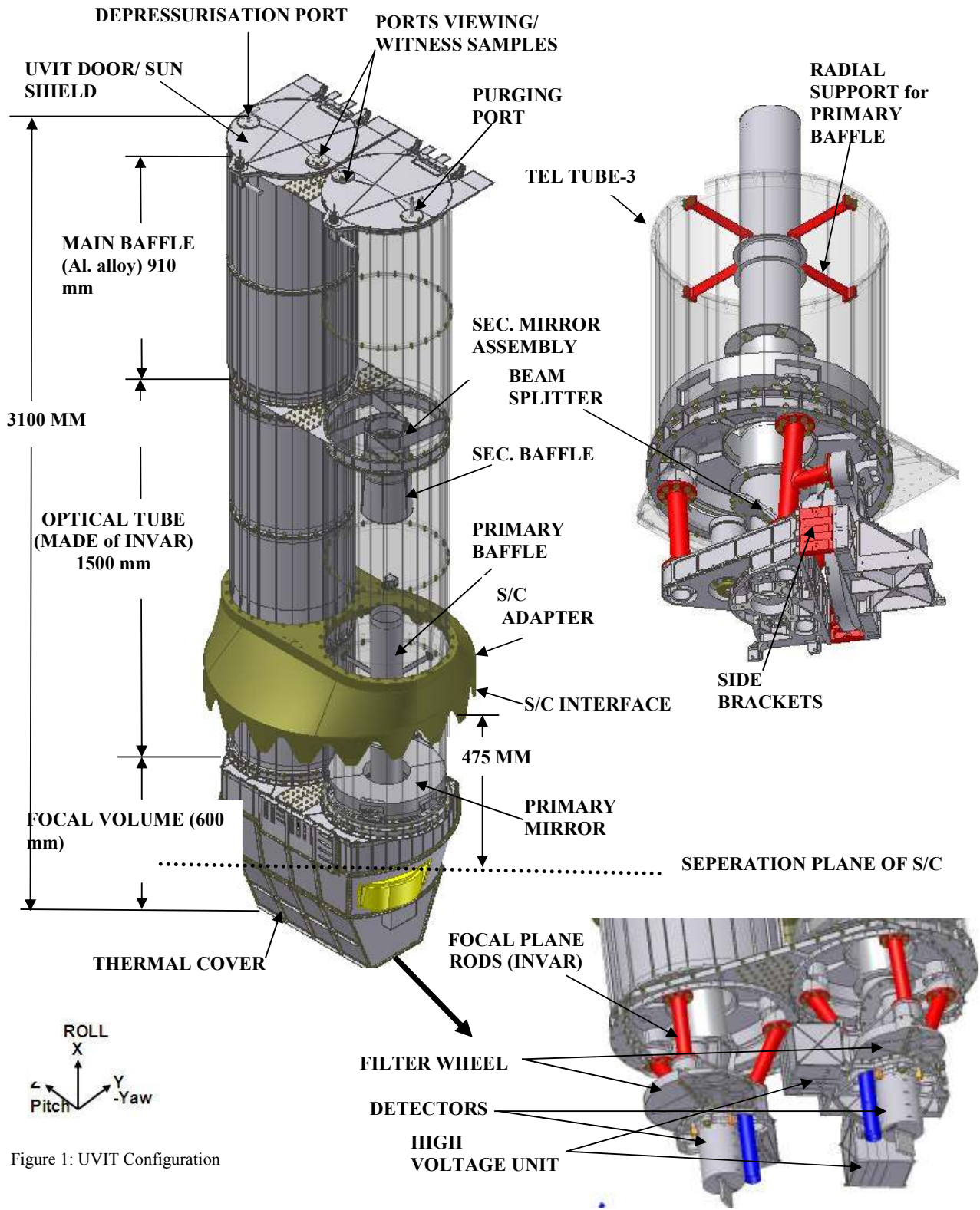

Figure 1: UVIT Configuration

The overall specifications of the payload are presented in Table 1, and details of the sub-systems are presented in the following sub-sections.

Table 1 Specifications of Ultra Violet Imaging Telescope

|  | **FUV** | **NUV** | **VIS** |
|---|---|---|---|
| Detector | Intensified CMOS Photon Counting/ Integration | Intensified CMOS Photon Counting/ Integration | Intensified CMOS Photon Counting/ Integration |
| CMOS chip | Fillfactory/ Cypress STAR250, 512x512, 25µm pixels | | |
| Telescope Optics | Ritchey-Chertian 2 mirror System | Ritchey-Chertian 2 mirror System | Ritchey-Chertian 2 mirror System |
| Bandwidth | 130-180 nm | 200-300 nm | 320-550nm |
| Geometric Area (cm$^2$) | ~880 | ~880 | ~880 |
| Effective Area (cm$^2$) | > ~15 at peak | > ~50 at peak | > ~50 at peak |
| Field of View | ~28' | ~28' | ~28' |
| Spectral Resolution | <1000 A (depends on Choice of Filters) | <1000 A (depends on Choice of Filters) | <1000 A (depends on Choice of Filters) |
| Spatial Resolution | <1.8 arc second | <1.8 arc second | <1.8 arc second |
| Time Resolution | <10 ms (for Partial field) | <10 ms (for Partial field) | <10 ms (for Partial field) |
| Typical Observation time per target | 30 min | 30 min | 30 min |
| Sensitivity (obs.time) | >20$^{th}$ Magnitude (5σ) in 200 s | - | - |
| Photometry Accuracy | 10% | | |
| Total Mass (Kg) | 230 Kg | | |
| Total Power (watts) | 85 watts (peak 117 watts) | | |
| Sun-avoidance angle | 45 deg | | |

## 3.2 Optical Design

Each of the UVIT telescope is based on Ritchey-Chretien configuration with an aperture of ~375mm and a focal length of ~ 4750 mm. Figure 2 and figure 3 shows the optical layout of the FUV and NUV-VIS telescope respectively.

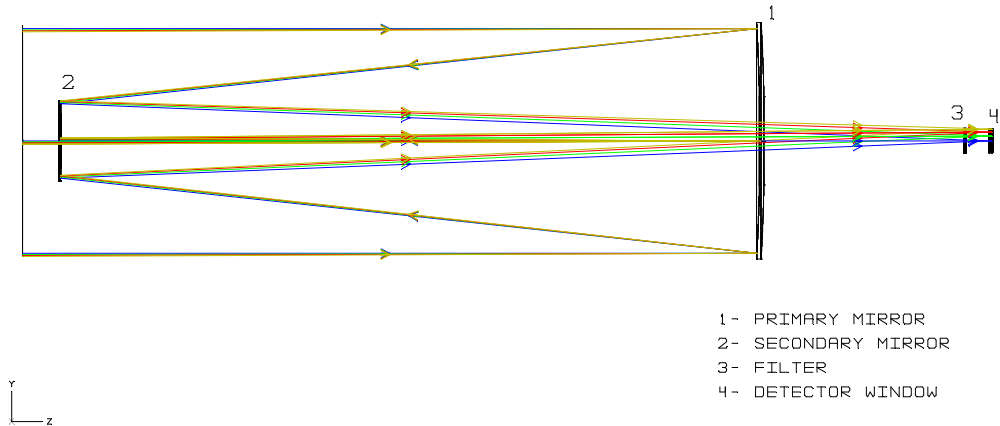

Figure 2 Optics of FUV Telescope; the position marked 'filter' carries a filter wheel with selection of 5 filters, 2 gratings & a block.

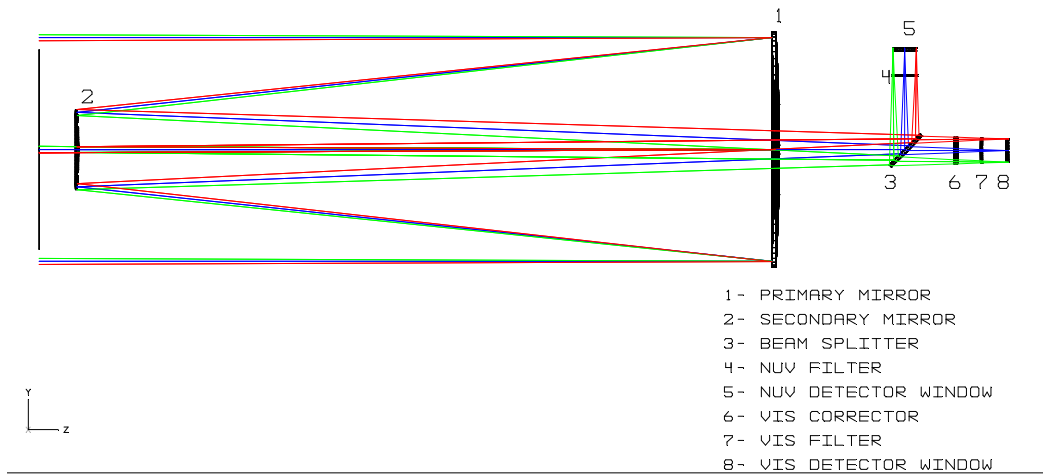

Figure 3 Optics of NUV-VIS telescope; the position marked as 'filter' carry a filter wheel; in NUV channel wheel has a selection of 6 filters and a grating and a block, while in VIS channel wheel has a selection of 5 filters and a block.

In both telescopes, primary mirror is a solid mirror with diameter of 375mm and a central hole of 155mm. Mirror material used is zerodur with the surface error of better that $\lambda/50$ rms and micro roughness better that 15Å rms. The surface of primary is concave on-axis hyperboloid with radius of curvature 3541mm and conic -1.129. Primary mirror is mounted in telescope by side mounts; i.e 3 bipods @120° apart and the mounts are glued to mirror and mounts to be fixed on a ring.

Secondary mirror for both the telescopes is also a solid mirror made of Zerodur material. Surface of secondary mirror is convex on-axis hyperboloid with radius of curvature 1867mm and conic -6.3565. Its diameter is ~140mm and it is mounted using 3 blade cell mount and mount is glued to the mirror cylinder rim. Surface error and micro roughness of secondary mirror are also same as for the primary mirror.

Average reflectivity of both the mirrors is maintained identical; i.e better than 60% for the wavelength band of FUV (130nm-180nm), better than 70% for the wavelength band of NUV (180nm-200nm) and better than 80% for VIS (200nm-600nm).

Appropriate baffling is provided at necessary places in the optical design. To avoid any cosmic and bright light hitting the detector, in each telescope there are 2 main baffles above telescope tubes, one primary baffle near primary mirror and one secondary baffle near secondary mirror as shown in figure 1.

Filters for each channel are mounted on a wheel. FUV channel has 5 filters, 2 gratings and 1 block, totaling 8 slots on wheel. NUV channel has 6 filters, 1 grating and 1 block totaling 8 slots on wheel whereas VIS has 5 filters and 1 block totaling 6 slots on the wheel. Table 2 lists the name of filter used for each channel.

Table 2 List of Filers & gratings used in UVIT

| S.No. | Filter Type | Filter Thickness | Passband | Material |
|---|---|---|---|---|
| **FUV CHANNEL** | | | | |
| 1. | Block with Aluminium | | | |
| 2. | Calcium Fluoride - 1 | 2.50 mm | >125 nm | |
| 3. | Barium Fluoride | 2.40 mm | >135 nm | |
| 4. | Sapphire Window | 2.00 mm | >142 nm | |
| 5. | Grating - 1 | 4.48 mm | | |
| 6. | Silica | 2.70mm | > 159mm | |
| 7. | Grating – 2 | 4.48 mm | | |
| 8. | Calcium Fluoride – 2 | 2.50mm | >125mm | |
| **NUV CHANNEL** | | | | |
| 1. | Block with Aluminium | | | |
| 2. | Fused Silica Window | 3.00 mm | > 159nm | |
| 3. | NUVB15 | 2.97 mm | 200 nm – 230 nm | Silica (UV) |
| 4. | NUVB13 | 3.15 mm | 230 nm – 260 nm | Silica (UV) |
| 5. | Grating | 4.48 mm | | |
| 6. | NUVB4 | 3.33 mm | 250 nm – 280 nm | Silica (UV) |
| 7. | NUVN2 | 3.38 mm | 275 nm – 285 nm | Silica (UV) |
| 8. | Fused Silica Window | 3.30mm | > 159nm | |
| **VIS CHANNEL** | | | | |
| 1. | Block with Aluminium | | | |
| 2. | VIS 3 | 3.00 mm | 400 nm – 530 nm | UBK7 |
| 3. | VIS 2 | 3.00 mm | 370 nm – 410 nm | UBK7 |
| 4. | VIS 1 | 3.00 mm | 320nm – 360 nm | UBK7 |
| 5. | Neutral Density Filter | 3.00 mm | | |
| 6. | BK7 Window | 3.00mm | | |

### 3.3 Mechanical and Thermal Design

The mechanical configuration of the payload is described and shown in the Figure 1 and its subsystems interfaces and its specifications are presented. The primary and secondary mirror system separation is maintained by invar tubular structure (~1500 mm), made in 3 segments. The Telescope ring (TR ring) at bottom supports primary mirror, the secondary is supported at top end by spider ring (SPDR). The TR ring also supports focal volume elements by a system

3 invar rods (FR's), bottom segment of the telescope tube (TT3) and thermal cover. The SPDR is 4 blades tangential system holding secondary mirror at center, the provision for required tilt and de-center are given at the secondary mirror interface.

Over the metering structure is a main baffle section (~910 mm), which also provides interface to the door, which, when deployed in space also serves as sun shield. Below the metering structure is focal volume, where in all detectors, filter wheel, drive system and high voltage units are contained. The detector bracket supported at the bottom end of the FR's serves a supporting structure for the above sub-systems. The entire focal volume is covered by an Al. sheet metal enclosure (thermal cover). This design is modified from its early version[4].

The titanium satellite adapter on the UVIT structure provides the interface between UVIT and spacecraft (ASTROSAT). The attachment is through 18 nos tabs on the titanium adapter, which are held against satellite cylinder by M6 bolts, which get engaged in to the plate nuts riveted in the tabs. The satellite cylinder is provided with two cut outs to take the cable harness in to the spacecraft. The titanium satellite adapter also has a master ref. cube fitted on it, to serve as a ref while integrating with spacecraft. ICDS stating all the above requirements and are also met.

The structural materials and its properties used in payload are detailed in table 3. As the telescope demands high degree of dimensional stability under changing temperatures, the telescope tube which holds the primary and secondary mirror is made of Invar36 material. The Telescope tube made of Invar36 material is made in 3 parts as tube1, tube2 and tube3 and three tubes are connected together by Titanium bolts. These tubes holds the Primary and secondary mirror to at required position with high degree of positional accuracy. Other parts of the telescope like the baffles and detector mounts are made of aluminum alloy material which has high tensile strength and light in weight. Aluminum 6061T6 / IS: 64430 alloy has been used for making above said components. Two telescopes have been coupled to an adapter and further the adapter with the telescopes is mounted to the satellite cylinder. Considering the high stiffness, light weight and its thermal properties, the adapter is made of Titanium alloy material, Ti6Al4V-Grade5. The adapter is of unique shape with diameter on 877 mm with height of 284 mm has been scooped out from a titanium forged block.

The mass of UVIT has two parts, UVIT mass attached to central cylinder of satellite is 202 Kg, and UVIT electronics packages mass on satellite equipment panels is about 28 Kg amounting the total UVIT mass as 230 Kg.

Table 3 Material Used and their properties

| Name of Material | Youngs modulus (Gpa) | Density (kg/m$^3$) | Poisons ratio | Tensile yield strength (Mpa) | CTE (µm/m $^0$C) |
|---|---|---|---|---|---|
| INVAR | 148 | 8005 | 0.29 | 240 | 1.3 |
| Al. Alloy (6061T6) | 68.9 | 2700 | 0.33 | 320 | 23.6 |
| Titanium alloy (Ti6AL4v) | 115 | 4300 | 0.3 | 880 | 8.6 |

The payload is modeled completely with a finite element model, Static, modal and frequency response analysis were carried out for sine and random inputs. The quasi-static design margins, bolt forces (subsystem interfaces), first resonant frequency, and dynamic responses at the interfaces were analyzed. Buckling analysis was also carryied out to find the lowest buckling factor.

Thermal design and analysis is concerned with predicting temperatures of the payload in a specific orbital thermal environment. The numerical thermal model is used as the working tool in the development of the satellite thermal control system. It is used to predict temperature on a large scale, with most structures and other components interacting with one another and with the surrounding environment. The mode of heat transfer in the payload system is generally through conduction and radiation heat transfer except in the case at the launch pad. The ambient temperature and heat loads influences overall temperature distribution in the payload. Thermal control system of the payload employs a passive method of isolation by multi layer insulation (MLI), and heaters under closed loop control. The UVIT is wrapped all over with MLI. Heaters and Optical solar reflector (OSR) are used to maintain the temperature. Thermal load are the internal and external loads. Internal loads due to internal power dissipation of the filter motor, detector and high voltage box, which are located in the focal volume unit. External loads are due to the sun, albedo of the earth and earthshine and are evaluated based on the orbit parameters. Following table 4 shows the thermal specifications and achieved values as per thermal model design. Out of all the requirements for the payload, the most difficult to control is the variation in relative alignment of the two telescopes, as a variation of 1" over a typical exposure of 1000 s can lead to an additional blur of ~ 0.3" rms in the image size (due to a less than perfect correlation between aspects of the two telescope during the period of integration). Relative alignment of axis of the two telescopes should not drift by more than 0.5 arcsec in any 15 minutes (a typical period of observing a field in UV ) of the orbit; but it can change by 30 arcsec on long time scales subject to a drift rate less than 1 arcsec per 15 minutes.

Table 4 Thermal Requirements and achieved thermal model results

| Requirements | Achieved values |
|---|---|
| Temperature of telescope tubes to be between 18°C and 22°C | 17.5°C(Minimum) and 22.8°C(Maximum) in cold invar case |
| Axial variation of temperature on telescope tubes to be within +/-2°C | 2.3°C on NUV side in cold focal case and cold invar cases. |
| Circumferential variation of temperature on telescope tubes to be within 5°C | 2.8°C in cold focal case. |
| Temporal variation of temperature at a given point within 1000secs (~15 minutes) (in quasi steady state) to be within 0.3°C | 0.77°C (TT2 bottom portion in FUV side) in hot focal case (Maximum) and 0.02°C (TT2 top portion in NUV side) in hot focal case (Minimum) |
| Temperature of elements in the focal plane volume to be between 15°C and 20°C | 12.7°C (Minimum) 20.6°C (Maximum) |
| Temperature (during operation) of detectors (CPU's) between 0°C and 20°C | 16.4°C (Minimum) 17.9°C (Maximum) |
| Temperature (during operation) of High Voltage Units (HVU's) between 0°C and 30°C | 12.7°C (Minimum) 18.4°C (Maximum) |
| Duty cycle of heaters not to exceed 65% | 64% in MB1 in cold invar case. |

### 3.4 Detector and Electronics

The detector system for each of the 3 channels is generically identical. Each consists of an image intensifier tube, fiber optic coupling and an imaging sensor. A typical photon event undergoes the following: in the intensifier tube, the incident UV/optical photon ejects a primary electron from the Photo-Cathode (PC), which is accelerated towards and multiplied within the Micro-Channel-Plate (MCP). Gain of MCP is of order $10^7$. The large number of secondary electrons emerging from the MCP are further accelerated towards the Anode (A) and which is coated with a phosphor. The light emission of the phosphor is topologically mapped (with a reduction of ~ 3:1) onto the Fillfactory/ Cypress STAR250 (512x512 pixel on 25μm pitch) [5] CMOS image sensor by the fiber-optic taper. The intensifier tube requires 3 sets of high voltages for its Photo-Cathode (V_PC), Micro-Channel-Plate assembly (V_MCP) and the Anode (V_A) for its operations. The V_PC can be used to electronically switch the channel ON/OFF, and V_MCP is for controlling the 'gain' (=amount of light signal from the phosphor of the intensifier tube per primary photon event). For each intensifier tube, a set of optimum values for these 3 voltages are available based on ground tests & calibrations.

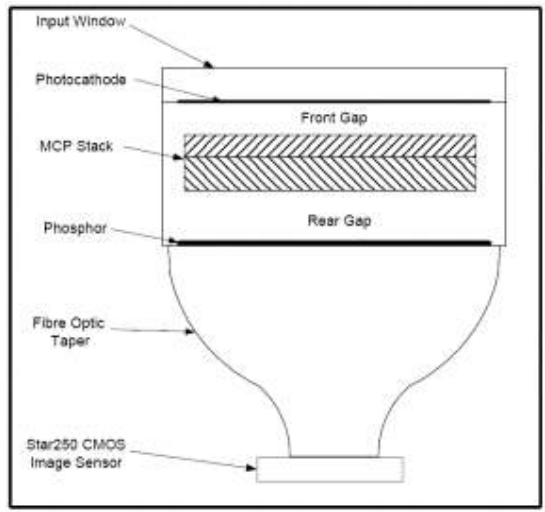

Figure 4 Photon Counting Detector

The full circular field-of-view (28 arc-min diameter) fits into the square image sensor (512x512 pixels) which is read at a selectable frame-rate up to a maximum of ~ 29 frames/sec. Either the entire array or a rectangular 'window' of selectable dimensions and position can be read out; the maximum frame-rate increases in inverse proportion to area of the selected window  Each of the 3 detectors can either be used in the 'Photon Counting' mode or 'Integration' mode. In the 'Photon Counting' mode, the photon events are detected and their centroids are calculated onboard. In the 'Integration' mode, the raw frames are sent to ground. Normally, the FUV & NUV channels are used in 'Photon Counting' mode, and the VIS channel in 'Integration' mode. The high voltage supply of each of the detector is mounted near the detector.  Electronics for control and processing of the signals from of the detectors is placed inside a box which is mounted on deck of the S/C; the same electronics provided interface with the S/C for telemetry and telecommand operations. For details of tests see reference 6, 7 & 8.

The UVIT Filter system (for each of the 3 channels) consists of two distinct physical sub-units: Filter-Wheel-Motor Assembly which carries the filter wheel and Filter-Wheel Drive Electronics (FWDE) which drives the motor. The Filter Wheel Motor Assembly is located such that the filters are ~ 40 mm in front of the detector-window. All the three Filter Wheel Drive Units are located in a box mounted on deck of the S/C.

All the three channels (FUV, NUV & VIS) of UVIT individually commanded by spacecraft on-board controller and payload health parameters (voltages, currents & temperatures) will be received by spacecraft on-board controller. These health parameters will be transmitted by the spacecraft to ground station for reviewing and recording purpose.

**3.5  Contamination Control for UVIT**

The absorption cross-sections in ultraviolet are very large, and great care needs to be taken to avoid any contamination of the optical surfaces of the UVIT, on which the radiation falls, during any stage of fabrication/testing/assembly or packing/transport etc. Therefore, great care is taken in selecting materials which go in the optical cavity. This section describes the possible ways to minimize the chances of contamination. The contamination of an optical surface can be either due to deposition of the particles or due to deposition of the molecules. Contamination due to particles can be minimized by insisting that the optical surfaces are only opened in areas with clean air; for the purpose of the window of CPUs, Mirrors, etc., it is best that these are only opened in Class 100 air, and this exposure is minimized. The control of molecular contamination is far more difficult to achieve. In this note we are primarily concerned with control and monitoring of the molecular contamination.

All materials and process used in any part of the instrument are subjected to a screening process and based on contamination clearance the material or process are added in to the approved Declared Material List(DML)Declared Process List(/DPL). Only those materials are selected for further checks which have TML < 1% and CVCM < 0.1%. The materials passing these criteria are further tested for their potential for contamination before being accepted for use.

For UVIT 20mm diameter x 2mm thick windows of MgF2 are used as witness samples for measuring possible contamination due to the material. The window is exposed to the material kept in vacuum at a high temperature for a long period, e.g. at 80-120 deg C for 24 -72 hrs. The material degassed from contaminant deposits over the witness sample which is kept at lower temperature (~30 deg C). The transmission of the window in UV range (120nm to 180nm) is measured and compared with its original transmission. The degradation in transmission gives the measure of potential of contamination from the particular material.

In order to monitor that the optical surface has not been contaminated during tests/assembly/transport etc. a witness sample—either a MgF2 window or a mirror with a layer of MgF2-- always accompanies the optical component. The witness sample is periodically checked for its transmission/reflection in 120-180 nm range, and any reduction in the transmission/reflection is indication of contamination of the component. The key specification is that the overall transmission/reflection loss on a witness sample, during all the operations from assembly to launch, be < 5%. This affects the effective area of the telescopes, and a loss of 5% implies an overall loss of 15-20% in the effective area.
So far the witness windows kept with the detectors and the mirrors have not shown any measurable loss of transmission.

### 3.6 Spatial resolution and Drift of S/C

As mentioned earlier, pointing of the S/C is not stable for the required spatial resolution of 1.8". In order to avoid blurring short exposures (e.g. < 1 s) are taken and all the exposures are combined through shift and add process to get an effective long exposure image. Typically, the number of ultraviolet photons is too small to give a sensible image in an exposure of 1 s. However, the visible channel can collect enough photons in 1 s to make a useful image. Therefore, images from the visible channel are obtained every ~ 1 s to estimate relative drift of the S/C during a long exposure. This process has been simulated for realistic values of the parameters and is seen to work[9]. This process works well as long as the relative aspect of the three channels does not change over period of the long exposure, and the mechanical structure is made of Invar 36 to minimize any drifts in the relative aspects due to thermal effects.

## 4. CURRENT STATUS

After successful testing of the engineering model, the flight model is being assembled at MGKM Space Science Laboratory of IIA, see figure 5. The telescope for NUV/VIS channels has been fully assembled and tested to give images with FWHM, 1.5". The FUV telescope is under assembly and would be tested in the month of June. The payload is expected to be delivered to ISRO in the month of July, for environmental testing and integration with the S/C. The expected launch of the mission is second quarter of 2013.

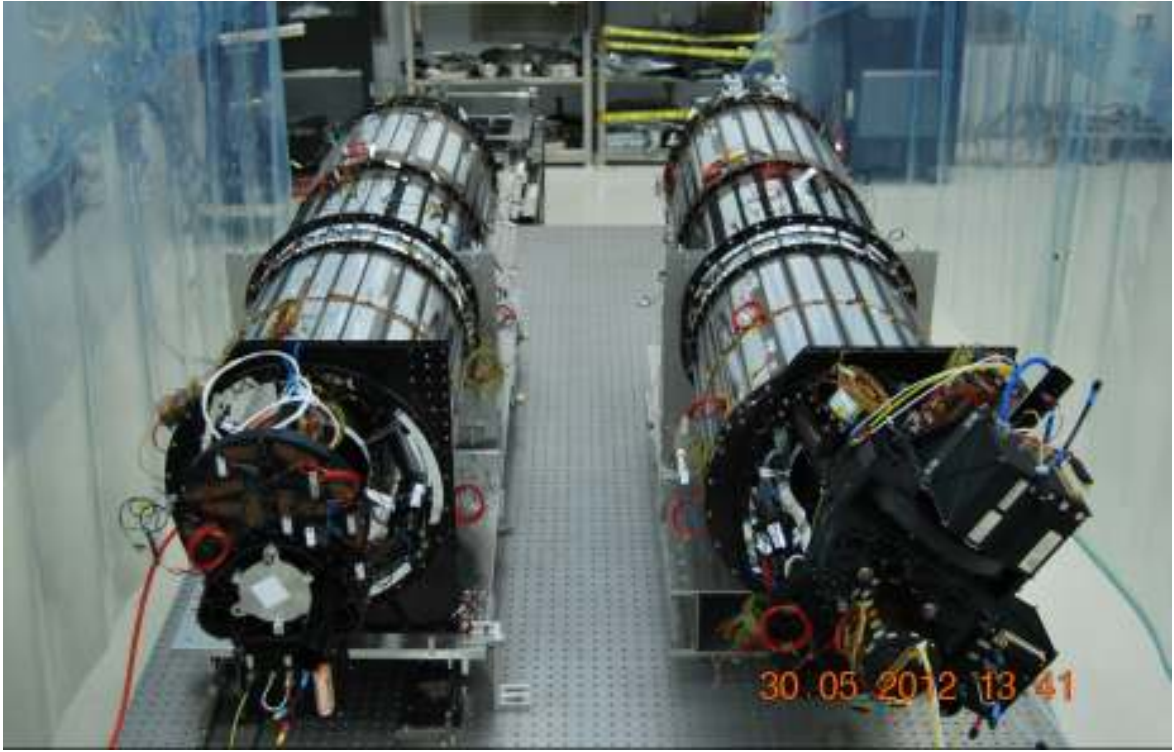

Figure 5 Assembled UVIT FUV and NUV-VIS Flight Model telescope

## 5. CONCLUSION

The design and its implementation, has been described for the Ultra Violet Imaging Telescope for the multi-wavelength Indian mission ASTROSAT. The results of the tests on components and the assembled telescopes show that the key specifications of spatial resolution < 1.8" FWHM and a sensitivity of mag 20 in 200 s for the FUV channel are expected to be met. The payload is expected to be launched in the early part of year 2013.

## ACKNOWLEDGMENTS

The UVIT project is collaboration between the following institutes from India: Indian Institute of Astrophysics (IIA), Bengaluru, Inter University Centre for Astronomy and Astrophysics (IUCAA), Pune, and National Centre for Radioastrophysics (NCRA) (TIFR), Pune, and the Canadian Space Agency (CSA). The detector systems are provided by the Canadian Space Agency. The mirrors are provided by LEOS, ISRO, Bengaluru and the filter-wheels drives are provided by IISU, ISRO, Trivandrum. Many departments from ISAC, ISRO, Bengaluru have provided direct support in design and implementation of the various sub-systems.